\documentclass{emulateapj}
\usepackage{color}
\usepackage{multirow}
\usepackage{units}
\usepackage[colorlinks,linkcolor=blue,anchorcolor=green,citecolor=blue]{hyperref}
\shorttitle{Constraining the ejecta of GW170817}
\shortauthors{Huang \& Li}

\begin{document}

\title{Constraining the ejecta for the nonthermal emission from GW170817}
\author{Yan Huang and Zhuo Li}
\affil{Department of Astronomy, School of Physics, Peking University, Beijing 100871, China; hyan623@pku.edu.cn\\
Kavli Institute for Astronomy and Astrophysics, Peking University, Beijing 100871, China
}
\begin{abstract}
We consider a simple model for the nonthermal emission from GW170817, in which a quasi-spherical ejecta is released in the merger event, with the kinetic energy distributed over the momentum as $E(>\gamma\beta)\propto(\gamma\beta)^{-k}$. The ejecta drives a shock into the medium and gives rise to synchrotron radiation. Using multi-band observations we constrain that $k\approx6.7$;  (assuming medium density of $\sim10^{-2}\rm cm^{-3}$ and postshock magnetic field carries a fraction $10^{-5}-10^{-3}$ of the postshock internal energy) the total kinetic energy is $(0.3-5)\times10^{51}$erg; the slowest ejecta velocity is $\sim(0.7-0.8)c$; and the fastest ejecta has a Lorentz factor of $\sim4-7$. We conclude that the sub-relativistic dynamical ejecta responsible to the kilonova cannot produce the nonthermal emission. The co-existence of the nonthermal and thermal kilonova emission implies that two corresponding ejecta are ejected at different angles.
\end{abstract}


\section{Introduction}
The discovery of the gravitational wave event GW170817 \citep{gw} together with its electromagnetic counterparts marks the dawn of the multi-messenger astronomy era \citep{mm}. The properties of the gravitational wave signal imply that a neutron star merger occurs, which is further strengthened by the detected electromagnetic emission. There are both thermal and nonthermal emission associated with this event. The thermal emission in UV/optical and infrared band well reveals the predicted kilonova emission powered by the radioactive decay of heavy elements synthesized in the merger ejecta \citep[e.g.,][]{cou17, mm,pian17,arc17,evans17}. The nonthermal emission is detected across a wide frequency range. A gamma-ray burst (GRB) is detected $\sim1.7$ s after the gravitational wave event \citep{GBM,2017ApJ...848L..15S}. X-ray and radio emission starts to be detected after $\sim9$ days and $\sim16$ days, respectively \citep{2017Natur.551...71T,2017ApJ...848L..25H,2017ApJ...848L..20M,2017Sci...358.1579H,2017ApJ...848L..21A}, and continues brightening as $F_\nu\propto t^{0.8}$ until $\ga120$ days \citep{mooley17,margutti18,ruan17,radio-turn}. The X-ray and radio flux are consistent with a single-power-law spectrum, $F_\nu\propto\nu^{-0.6}$ \citep{mooley17}. The optical kilonova emission subsides after $\sim15$ days \citep{mm,pian17,arc17}, but the optical emission rises again after $\sim100$ days \citep{optical1,margutti18}, with a flux well consistent with the spectrum crossing X-ray and radio bands.

While the thermal component is widely believed to be produced by the non-relativistic (NR) kilonova ejecta, the origin of the nonthermal component is less understood, although very likely it arises from synchrotron radiation from a shock in the event. Some nonthermal emission models have been proposed, which can be roughly divided into two classes, one is the quasi-spherical explosion model \citep{cocoon,mooley17,x-decreasing,troja18} and the other is the off-axis jet model \citep{lazzjet,mur17,margutti18,optical1,x-decreasing,troja18,resmi18}. As already noted by many authors, the most simplified top-hat jet model cannot explain the late time X-ray and radio emission \citep[e.g.,][]{mooley17}, and a structured jet with angular dependent structure is needed. In quasi-spherical models, a simple homogeneous ejecta cannot work either, but a stratified ejecta with energy distribution over ejecta velocity can account for the observation \citep{mooley17}. However, the origin of the spherical ejecta is unclear, i.e., whether it can be the kilonova-related ejecta, or cocoon \citep{mooley17,dyn_ej,cocoon,nakar_whole}, or even other origins?

As pointed out by \cite{np18}, even in the structured jet model the dominant emission is that from wide angle ejecta approaching observers. Here we consider the spherical model, and constrain the physical parameters, e.g., the kinetic energy and the velocity, of the shock with observation, which may give hints to the origin of the ejecta for nonthermal emission.
\S2 describes the model of the ejecta-driven shock and its radiation, \S3 shows how we constrain model parameters with observations and the results, and finally \S4 is conclusion and discussion on the ejecta properties relevant to the nonthermal emission.

\section{Model}
Consider that a spherical ejecta is released from the GW170817 event and driving a shock wave into the medium. The kinetic energy of the ejecta is distributed over the ejecta momentum as $E(>\gamma\beta)=E_1(\gamma\beta)^{-k}$ \citep{mooley17} with $\gamma_{\min}<\gamma<\gamma_{\max}$, where $\gamma$ is the bulk Lorentz factor (LF) of ejecta, $\beta=(1-1/\gamma^2)^{1/2}$ is the bulk velocity in unit of $c$, and $E_1$ and $k$ are constants. Assume the medium is uniform with density $n$. This is reasonable since GW170817 lies at the outskirs of an early-type galaxy \citep{mm}, where clean environment is expected.

The interaction of the ejecta with the medium material generates a double shock structure - a reverse shock going into the ejecta and a forward shock sweeping up the medium material. We consider emission from the forward shock only\footnote{It is reasonable to neglect the emission from reverse shock, because the nonthermal spectrum appears to be a single power law from radio to X-ray bands, implying only one component, thus emission from reverse shock, if any, should be dominated by the forward shock one.}. By the shock's jump conditions we obtain, for the forward shock, the postshock energy density $U'$, particle density $n'$, and the shock's bulk Lorentz factor (LF) $\Gamma$ \citep{bm76},
\begin{eqnarray}
  \frac{U'}{n'}&=&\gamma m_pc^2,\\
  \frac{n'}{n}&=&\frac{\hat{\gamma}\gamma+1}{\hat{\gamma}-1},\\
  \Gamma^2&=&\frac{(\gamma+1)[\hat{\gamma}(\gamma-1)+1]^2}{\hat{\gamma}(2-\hat{\gamma})(\gamma-1)+2}.
\end{eqnarray}
Here $\gamma$ is the LF of the postshock fluid, and $\hat{\gamma}$ is the adiabatic index of the postshock material. We take $\hat{\gamma}=\frac{4\gamma+1}{3\gamma}$ \citep{dai99} so that $\hat{\gamma}$ is equal to $4/3$ and $5/3$ for ultra-relativistic and NR gas, respectively. The prime denotes that the quantity is measured in the comoving frame of the fluid. Note, these jump conditions are available for shocks with arbitrary values of $\gamma\beta$.

By energy density transformation, the postshock energy density (in the fixed frame) is given by $U=(U'+p')\gamma^2-p'$. The pressure (in comoving frame) is given by the equation of state $p'=(\hat{\gamma}-1)(U'-\rho'c^2)$. We adopt a thin-slab assumption for the shock by neglecting the structure behind the shock. Using the shock's jump condition, we then calculate the shock's energy, i.e., the total energy of the postshock material,
\begin{equation}\label{eq:energy}
  E_{\rm sh}=N_e\frac U{\gamma n'}=M_{\rm sh}c^2[f(\gamma)-1],
\end{equation}
where
\begin{equation}
  f(\gamma)=\frac{\gamma^3+(\hat{\gamma}-1)(\gamma^2-1)(\gamma-1)}\gamma,
\end{equation}
$N_e=(4/3)\pi R^3n$ is the total number of shocked baryons, with $R$ the shock's radius, and $M_{\rm sh}=N_em_p$ is the shock swept-up mass. It should be emphasized that eq. (\ref{eq:energy}) is available for strong shocks with arbitrary $\gamma\beta$, no mater ultra-relativistic ($\gamma\gg1$), mildly-relativistic ($\gamma\beta\sim1$), and Newtonian-phase ($\beta\ll1$) shocks.

For a shock with the LF of the postshock fluid being $\gamma$, the shock energy is provided by the kinetic energy of ejecta with bulk LF $>\gamma\beta$.
%{\bf From the observation, the nonthermal X-ray, optical and radio emission are consistent with the synchrotron radiation of the forward shock. So here, we only consider the forward shock emission, although %there would be a reverse shock traveling into the ejecta that could produce non-thermal emission as well. When we calculate the dynamics of the shock, We assume that the ejecta provide it's total kinetic %energy to the forward shock, by equating $E(>\gamma\beta)=E_{\rm sh}(\gamma)$. It should be noted that in a more realistic model, this should roughly half of total kinetic energy of the ejecta transfer to %the forward shock, but it has little effect on the dynamics of forward shock.
By equating $E(>\gamma\beta)=E_{\rm sh}(\gamma)$\footnote{Indeed the ejecta energy is shared by both the reverse and forward shocks, each roughly sharing half in thin-slab assumption. So we have neglect a factor of $\sim1/2$ in the l.h.s. of the equation, which has little effect on the result. }, we obtain the forward shock's dynamics, i.e., the $\gamma$ and $R$ relation,
\begin{equation}\label{eq:dyn}
  M_{\rm sh}(R)c^2=\frac{E_1(\gamma\beta)^{-k}}{f(\gamma)-1}.
\end{equation}
At observer's time $t$ the shock's radius is
\begin{equation}\label{eq:R-t}
  R=(1+\beta_\Gamma)\beta_\Gamma\Gamma^2ct,
\end{equation}
where $\beta_\Gamma=(1-1/\Gamma^2)^{1/2}$. Combining Eqs. (\ref{eq:dyn}) and (\ref{eq:R-t}) we can solve $\gamma(t)$ and $R(t)$ as function of $t$.

The shock can accelerate the swept-up electrons, compress and amplify the ambient magnetic field, thus give rise to synchrotron and inverse-Compton (IC) radiation. The energy density of the postshock magnetic field can be parameterized as a fraction $\epsilon_B$ of the postshock internal energy, $U_B'=B'^2/8\pi=\epsilon_B(U'-\rho'c^2)$, i.e., using the jump condition, the postshock magnetic field is given by
\begin{equation}
  B'=\sqrt{8\pi\epsilon_B\frac{\hat{\gamma}\gamma+1}{\hat{\gamma}-1}(\gamma-1)nm_pc^2}.
\end{equation}
The postshock electrons are expected to be accelerated to follow a power law distribution over the electron LF, $dn_e/d\gamma_e=C_e\gamma_e^{-p}$ at $\gamma_e\geq\gamma_m$. Denote $\epsilon_e$ the fraction of the postshock energy converted to electrons, then the minimum LF is given by
\begin{equation}
  \gamma_m=\frac{p-2}{p-1}\epsilon_e(\gamma-1)\frac{m_p}{m_e}.
\end{equation}

The cooling of electrons due to synchrotron and IC radiation is faster for larger electron LF, thus the distribution of electrons in high energy end may deviate from the power law of index $p$ due to fast cooling. Derive the electrons' cooling LF, $\gamma_c$, by equating the cooling time (in the fixed frame) $t_c(\gamma_e)=\frac{3m_ec}{4\sigma_TU_B'(1+Y)(1+\beta)\gamma \gamma_e}$ to the dynamical time $t$, we have
\begin{equation}
  \gamma_c=\frac{3m_ec}{4\sigma_TU_B'(1+Y)(1+\beta)\gamma t},
\end{equation}
where $Y$ is the Compton parameter of electrons, i.e. the ratio of electron's inverse-Compton power to synchrotron power.

A single electron of LF $\gamma_e$ produces synchrotron photons at characteristic frequency $\nu'(\gamma_e)=\frac3{4\pi}x\frac{\gamma_e^2eB'}{m_ec}$ with $x=0.23$ \citep{wijers99}, and the specific synchrotron power at $\nu'$ is $P_m=\frac{\sqrt{3}e^3B'}{m_ec^2}$, independent of $\gamma_e$. The characteristic frequencies emitted by electrons of LF $\gamma_m$ and $\gamma_c$ are, respectively,
\begin{equation}
  \nu_m=\frac43\gamma\nu'(\gamma_m)=\frac x{\pi}\gamma\frac{\gamma_m^2eB'}{m_ec},
\end{equation}
and
\begin{equation}
  \nu_c=\frac43\gamma\nu'(\gamma_c)=\frac x{\pi}\gamma\frac{\gamma_c^2eB'}{m_ec}.
\end{equation}

The synchrotron spectrum from power law distributed electrons can be approximated as a broken power law. At the frequency range above the synchrotron absorbtion frequency $\nu_a$ and $\nu_m$ but below $\nu_c$, the synchrotron flux is given by
\begin{equation}
  F_\nu=F_m(\frac{\nu}{\nu_m})^{-(p-1)/2}
\end{equation}
where the peak flux is $F_m=\frac{N_e\gamma P_m}{4\pi d_L^2}$ \citep{sari98}, with $d_L$ the luminosity distance.

\section{Constraints by observations}
The model parameters include: $E_1$, $k$, $n$, $\epsilon_e$, $\epsilon_B$, and $p$. We will constrain the parameters with observations of GW170817. The observational data considered include the radio data taken from \cite{2017Sci...358.1579H,mooley17}; the X-ray data (0.3 - 8keV) from \cite{latestX, nyn18}; and the optical data from \cite{optical1,margutti18}.
%i.e., $m = 26.44\pm0.14$mag at $t=110$ day, and $m = 26.90\pm0.25$mag at $t=137$ day. ACS/F606W provided by the HST team.(central wavelengths 589nm)
We take the luminosity distance $d_L=40\unit{Mpc}$ \citep{gw,hjorth17}.

\subsection{Electron index}
The observed spectra from radio to X-ray and from $\sim10$ days up to $\sim150$ days are well consistent with a spectral slope of $F_\nu\propto\nu^{-0.6}$ \citep{mooley17}. Moreover, the photon index of observed X-ray spectra are well fit to be $\Gamma_{ph}\approx1.6$ \citep{margutti18}, consistent with the radio to X-ray spectral index.  By synchrotron radiation model, this constrains the electron index, $p=2(\Gamma_{ph}-1)+1\approx2.2$. This is a typical value expected in diffusive shock acceleration process in relativistic shocks \citep{pvalue1,pvalue2,pvalue3,pvalue4}, strengthening a shock origin of the nonthermal emission.

\subsection{Ejecta profile}
\begin{figure}
\vskip -0.0 true cm
\centering
\includegraphics[scale=0.42]{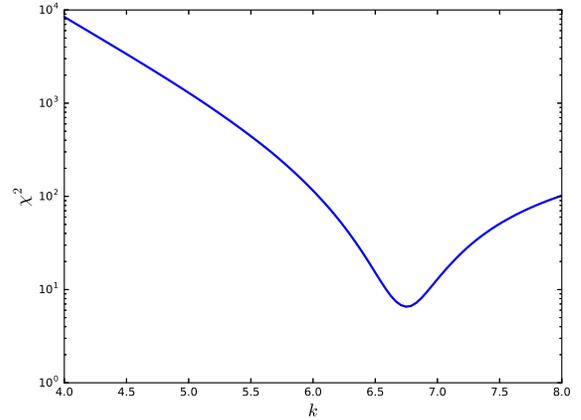}
\caption{$\chi^2$ as function of $k$ from the fitting of the 3-GHz light curve. The other parameter values are: $\epsilon_e=0.1$, $\epsilon_B=0.01$, $n=5\times10^{-4}\unit{cm^{-3}}$, and $E_1=2.3\times10^{51}$erg. The minimum $\chi^2$ appears at $k=6.7$.}
\label{Fig:chi2}
\end{figure}

\begin{figure}
\vskip -0.0 true cm
\centering
\includegraphics[scale=0.42]{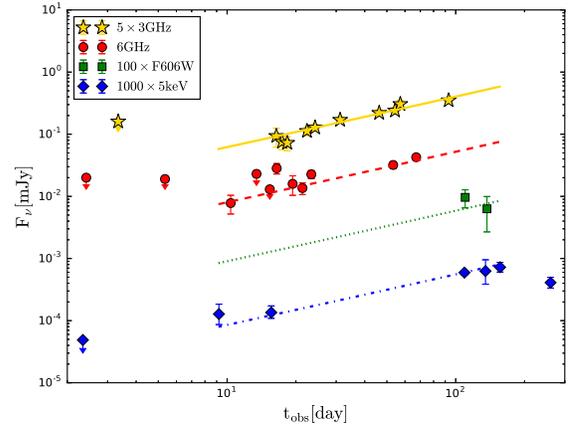}
\caption{The multi-band light curves of the model, with $p=2.2$, and $k=6.7$, in comparison with the multi-band observational data of GW170817. The other parameter values are as same as Fig \ref{Fig:chi2}.
%The other used parameter values in the model are $p=2.2$, $k=6.87${\bf use $k=7$!}, $\epsilon_e=0.1$, $\epsilon_B=0.01$, $n=10^{-4}\unit{cm^{-3}}$, and $E_1=2.72\times10^{52}$erg.
}
\label{Fig:light curve}
\end{figure}

The light curve slope mainly depends on $k$ value. Fixing $\epsilon_e=0.1$, $\epsilon_B=0.01$, and $n=5\times10^{-4}\unit{cm^{-3}}$, we let the other two parameters, $E_1$ and $k$, be free to fit the light curve data at $\nu=3$~GHz. When $k=6.7$ and $E_1=2.3\times10^{51}$erg, one obtains the minimum $\chi^2$ in the fitting. Fig \ref{Fig:chi2} shows $\chi^2$ as functions of $k$, with $E_1=2.3\times10^{51}$erg. If the parameter values of $\epsilon_e$, $\epsilon_B$ and $n$ are changed, the fitting results give a different value for $E_1$, but $k=6.7$ is almost unchanged. Using $p=2.2$ and $k=6.7$, the data can be well fit by the model, as shown in Fig \ref{Fig:light curve}. We only fit the data up to 156.4 day, around which there seems to be a light-curve turnover \citep{radio-turn, ale18, nyn18}, indicating that the slowest-moving material has caught up with the shock, and there is no significant energy injection into the shock any more. In the following we fix $k=6.7$ and constrain the remained parameter values.

\subsection{Energy normalization, medium density and magnetic field}
The data is not enough to constrain all the parameters, thus we fix $\epsilon_e=0.1$ in the following, because this is the typical value derived from observations of many kinds of shocks, e.g., NR supernova remnant shocks and relativistic GRB afterglow shocks. The remained main parameters are $E_1$, $n$ and $\epsilon_B$.
Here we will use the following two factors from observations to constrain them:
\begin{enumerate}
  \item $F_\nu = 0.048$~mJy at $\nu=3$ GHz and $t = 54.27$ days. This is from VLA observation \citep{mooley17}.
  \item $\nu_c>10^{18}$Hz at $t=156.4$ days. This is the latest X-ray observational epoch, where the photon index of the X-ray spectrum in 0.3-8keV is $\Gamma_{ph}\approx1.67$ \citep{latestX}. Nevertheless, the radio to X-ray spectrum at this time is also consistent with a single power law of $F_\nu\propto\nu^{-0.6}$. Thus both the X-ray spectrum and the radio to X-ray spectral slope imply that the cooling frequency $\nu_c$, assuming only synchrotron cooling ($Y=0$), should be beyond X-ray band.
 % \item $\nu_c>10^{18}$Hz at $t=15.4$ days. At this time the bright optical thermal kilonova emission still exists, so the electron cooling may be dominated by IC scattering the inner-coming kilonova % photons. We take $U_r'=U_{ph}'$. From observation, the bolometric luminosity at this time is $L_{bol}=1.23\times10^{40}$erg, and the temperature from the thermal spectrum is $T\approx2\times10^3$K.
\end{enumerate}

\begin{figure}
%\vskip -0.0 true cm
\centering
\includegraphics[scale=0.42]{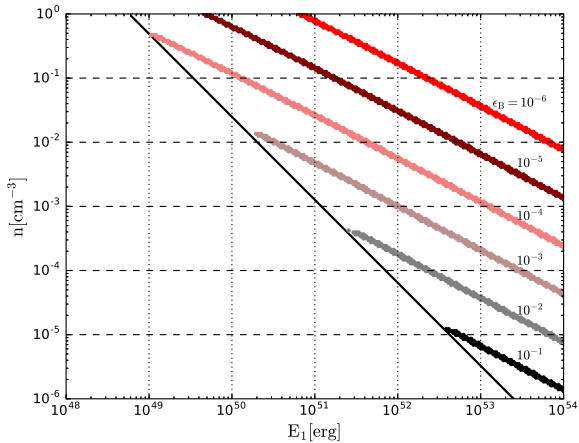}
\caption{Constraints of the values of parameters $E_1$, $n$ and $\epsilon_B$ with observations. The solid line is derived from combining factors 1 and 2. The allowed region in the $E_1-n$ space is that above the solid line. The shaded belts show the cases satisfying factor 1 by fixing $\epsilon_B$ as the marked values. The other parameters are $\epsilon_e=0.1$, $p=2.2$, and $k=6.7$.}
\label{Fig:energy-density}
\end{figure}

We are sampling the $(E_1,n,\epsilon_B)$ space to find the region that satisfy both factors 1 and 2. Indeed, combining these two factors, we have two equations for three unknown parameters. By setting $\epsilon_B$ as an variable, we can have a constraint of the other two on the $E_1-n$ space, as shown by the solid line in Fig. \ref{Fig:energy-density}.
%  \log n \geq -1.277\log E_1 +62.43.
The allowed region in $E_1-n$ space is that above the solid line.

The radio observations of short GRBs suggest that their average medium density is $n<0.15\unit{cm^{-3}}$ \citep{berger, fon15}. The observation of HI in the host galaxy of GW170817 gives a constraint of the medium density of $n<0.04\unit{cm^{-3}}$ \citep{2017Sci...358.1579H}. The location of GW170817 is observed to be at the outskirts of the host galaxy \citep{mm}, thus the medium density there should not be large, $n\ll1\unit{cm^{-3}}$. Taking $n<10^{-1}\unit{cm^{-3}}$ as an upper limit,  we can obtain, from Fig. \ref{Fig:energy-density}, a lower limit $E_1\ga4\times10^{49}$erg. For even lower density, a larger $E_1$ is required.

%If the ejected mass in the merger event is expected to be $<1M_\odot$ with a bulk velocity $\la0.3c$, one may assume $E_1<10^{52}$erg. If taking this upper limit to $E_1$, then Fig. \ref{Fig:energy-density} %shows that a lower limit $n\ga10^{-4}\unit{cm^{-3}}$ is obtained. ({\bf should be eliminated?})

In principle, we can also constrain $\epsilon_B$. In Fig \ref{Fig:energy-density} we show the sampled cases that satisfy factors 1 and 2 with $\epsilon_B$ being fixed (For given $\epsilon_B$ we require that the $\chi^2$ obtained in the fitting is $\chi^2<2\chi_{\min}^2$ where $\chi_{\min}^2$ is the minimum value). We see that a wide range of $\epsilon_B$ is allowed by the constraint, i.e., $10^{-7}\la\epsilon_B\la10^{-2}$. Note that the modeling of GRB afterglows usually give a wide range of $\epsilon_B$, but a lower limit is $\epsilon_B>10^{-6}$ \citep[see][and references there in]{lemoine13}.

\subsection{Ejecta kinetic energy and velocity}
\begin{figure}[h]
\vskip -0.0 true cm
\centering
\includegraphics[scale=0.42]{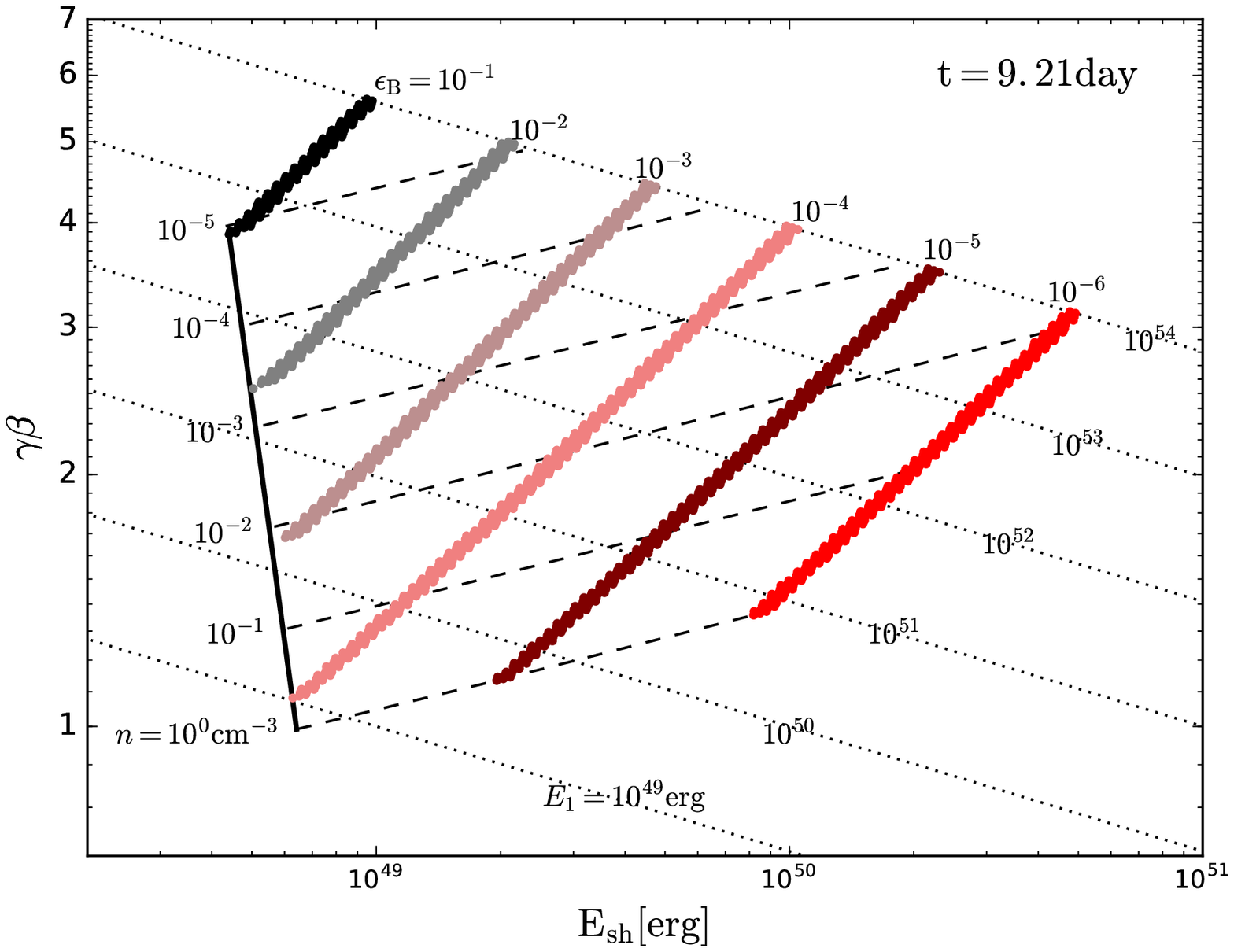}
\includegraphics[scale=0.42]{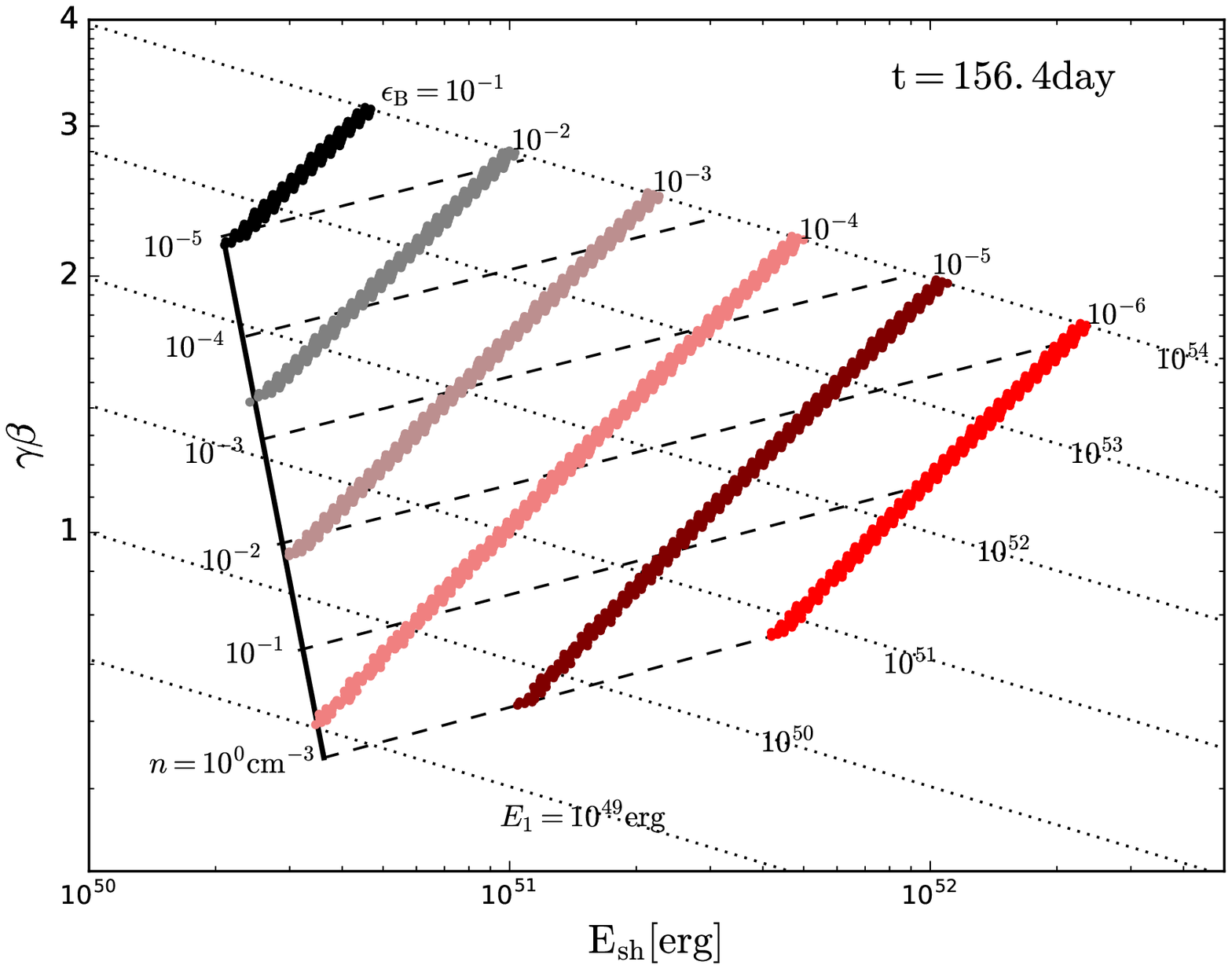}
\caption{Constraints of the total energy ($E_{\rm sh}$) and velocity ($\gamma\beta$) of the shock, at time $t=9.21$ (upper panel) and $156.4$ days (lower panel). The solid line corresponds to constraint by combining factors 1 and 2, and the allowed region is that above the solid line. The shaded belts correspond to the cases with fixed $\epsilon_B$ values as those in Fig \ref{Fig:energy-density}. The dotted lines show functions of $E_{\rm sh}=E_1(\gamma\beta)^{-6.7}$ with fixed $E_1$ values. The dashed lines show the $E_{\rm sh}-\gamma\beta$ relation (Eq. \ref{eq:dyn}) with fixed $n$ (and varying $E_1$). }
\label{Fig:energy}
\end{figure}

Given the values of $E_1$ and $n$, the dynamical evolution of the shock is determined by eqs. (\ref{eq:energy}), (\ref{eq:dyn}), and (\ref{eq:R-t}). The earliest detection of the non-thermal emission is at $t=9.21$ days, and the latest reported detection is X-ray detection at $t=156.4$ days. Both are consistent with the evolution with $p=2.2$ and $k=6.7$ (Fig \ref{Fig:light curve}). We derive the LF $\gamma$ and total energy of the shock $E_{\rm sh}$ at these two epoches for the allowed $E_1$ and $n$ values (Fig \ref{Fig:energy-density}). The results for $\gamma$ and $E_{\rm sh}$ are shown in Fig \ref{Fig:energy}.

During the observational period, the light curves of all bands keep rising. This is due to slower ejecta with larger kinetic energy catch up with the shock and provide their energy to the shock, i.e., $E(>\gamma)\approx E_{\rm sh}(\gamma(t))$. So the shock velocity is decreasing with time, but the shock energy is increasing with time. The determination of the shock velocity at the earliest time gives a lower limit to the velocity of the fastest part of the ejecta, while the shock energy at the latest time gives the lower limit to the total kinetic energy of the ejecta.

We can see in Fig \ref{Fig:energy} that if taking $n<10^{-1}\unit{cm^{-3}}$, the lower limit to the LF of the shock at $t=9.21$ days is $\gamma\beta\ga1.3$, %$\gamma\ga1.4$,
i.e., $\beta\ga0.8$. So this means the fastest ejecta has a velocity $v\ga0.8c$. For lower medium density, the fastest velocity should be even larger. % $n<0.01,\gamma\beta\ga1.7$, i.e., $v\ga0.9c$.
On the other hand, Fig \ref{Fig:energy} shows that the lower limit to the shock energy at $t=156.4$ days is $E_{\rm sh}\ga2\times10^{50}$erg, if assuming $\epsilon_B\la10^{-2}$, implying that the total ejecta kinetic energy is $E_{\rm ej}\ga2\times10^{50}$erg. This constraint is insensitive to $\epsilon_B$ and/or $E_1$.

\section{Conclusion and discussion}
We have used a simple spherical shock model to explain the nonthermal afterglow emission of GW170817. We consider only emission from forward shock. The shock dynamics derived is available to not only ultra-relativistic, and Newtonian shock, but also to mildly-relativistic shock. The model parameters, except for assuming $\epsilon_e=10^{-1}$, are constrained by multi-band observations, and the main results are:
\begin{itemize}
    \item The merger ejecta profile is with $k\approx6.7$, and $E_1\ga4\times10^{49}$erg for $n<10^{-1}\unit{cm^{-3}}$.
    \item The fastest ejecta has $\gamma_{\max}\beta_{\max}\ga1.3$, i.e., $\beta_{\max}\ga0.8$, for $n<10^{-1}\unit{cm^{-3}}$. The slowest ejecta has $\gamma_{\min}\beta_{\min}\ga0.7$, i.e., $\beta_{\min}\ga0.6$, for $n<10^{-1}\unit{cm^{-3}}$.
    \item The total ejecta kinetic energy is $E_{\rm ej}\ga2\times10^{50}$erg, if $\epsilon_B\la10^{-2}$.
\end{itemize}

Recently there are some works claiming that the rising light curve turns over around $t_{\rm turn}\sim150$ days \citep{radio-turn, ale18, nyn18}. This turnover implies that there is no more energy injection, and then the light curve starts to decline. So, the lower panel (for $t=156$ day) in Fig \ref{Fig:energy} shows us the constraint of the total kinetic energy of the ejecta. The HI observation of the host galaxy constrained that $n<0.04\unit{cm^{-3}}$ \citep{2017Sci...358.1579H}. However the location of GW170817 appears to be within the effective radius of the host galaxy \citep{levan17}, 
%($R_{\rm eff}\approx3$ kpc, with stellar mass of $\sim10^{11}M_\odot$) 
the medium density of GW 170817's location may not be very low. So, a reasonable circumburst medium density for GW170817 could be $n\sim10^{-2}\unit{cm^{-3}}$. Moreover,  GRB afterglow modeling usually gives a postshock magnetic field with $\epsilon_B\sim10^{-5}-10^{-3}$ \citep[e.g,][]{lemoine13, san14, bar14}. Taking $n\sim10^{-2}\unit{cm^{-3}}$ and $\epsilon_B\sim10^{-5}-10^{-3}$, we can obtain with the lower panel of Fig \ref{Fig:energy} that the total kinetic energy is $E_{\rm ej}\sim(0.3-5)\times10^{51}$erg, and that the slowest ejecta has $\gamma_{\min}\beta_{\min}\sim1-1.5$, i.e., $\beta_{\min}\sim0.7-0.8$. For the same range of $n$ and $\epsilon_B$ value, the upper panel of Fig \ref{Fig:energy} implies that the fastest ejecta has a velocity $\gamma_{\max}\beta_{\max}\ga1.8$, i.e., $\gamma_{\max}\ga2$.

The gamma-ray emission from GW170817 may be produced by the fastest part of the ejecta. Its energy $E_{\rm fast}\approx E_\gamma\simeq6\times10^{46}$erg \citep{mm} is many orders of magnitude smaller, compared with the constrained kinetic energy of the ejecta. For $n\sim10^{-2}\unit{cm^{-3}}$ and $\epsilon_B\sim10^{-5}-10^{-3}$, the lower panel of Fig \ref{Fig:energy} also show that $E_1\sim10^{50.5}-10^{52.5}$erg, with which we can estimate the velocity of the fastest ejecta, $\gamma_{\max}\beta_{\max}=(E_1/E_\gamma)^{1/k}\sim3.6-7.1$, i.e., $\gamma_{\max}\sim3.7-7.2$.

We summarize the constrained results in Table \ref{Tab:constraint}. The results imply that the merger event of GW170817 should release a mildly relativistic ejecta with the total kinetic energy of $\sim10^{51}$erg, and the bulk velocity $\ga0.7c$. This velocity is different from the kilonova-related ejecta. The UV/optical and infrared observations indicate that the ejecta accounting for the thermal kilonova emission should have subrelativistic velocities $0.1-0.3c$ and a mass of $\sim0.05M_\odot$ \citep{waxman17}, difference from the nonthermal emission related ejecta. In fact, for dynamical ejecta mass of $<0.1M_\odot$ and with a bulk velocity of $\la0.3c$, the total kinetic energy is $E<0.8\times10^{52}$erg. Using $k=6.7$, one has $E(>0.3)\approx E_1\times0.3^{-k}<E$, thus $E_1\la3\times10^{48}$ erg, which implies by Fig \ref{Fig:energy-density} that $n\ga $ a few $ \rm cm^{-3}$. This is in conflict with the constraint by \citet{2017Sci...358.1579H} that $n<0.04 \rm cm^{-3}$. So, it can be ruled out that velocity-structured material can both produce the observed nonthermal emission and be produced by dynamical ejecta.

On the other hand, if the thermal and nonthermal emitting ejecta exist the same time, one may expect that this is in conflict with a radio or X-ray light curver turnover at $t_{\rm turn}\sim150$day, since the slower and energetic kilonova-related ejecta would continue to inject energy into the shock, and prevent flux decreasing. Therefore, we expect that these two ejecta may be ejected at different angles, e.g., it is likely that the nonthermal emission related ejecta is more toward the axis of the binary orbit, and the kilonova-related ejecta more toward the orbital plane.

The mildly-relativistic ejecta with large bulk velocity, $v_{\min}\ga0.7c$, may not be the dynamical ejecta from the merger, nor ejected by the disk wind, but need more violent origin. It has been proposed that if a relativistic jet is launched, the cocoon due to the jet propagation in some material envelope may drive a shock that accounts for the nonthermal emission \citep{cocoon,nakar_whole}. However, it is also possible that even without a collimated relativistic jet, a wide angle ``fireball", e.g., generated by neutrino annihilation above the accretion disk \citep[e.g.,][]{eichler}, may form a quasi-spherical shock and give rise to the nonthermal emission.

\acknowledgements{The authors thank the referee for helpful comments. This work is supported by the NSFC (No. 11773003) and the 973 Program of China (No. 2014CB845800).}

%\appendix
%\renewcommand{\appendixname}{Appendix~\Alph{section}}
%\section{Derivation of absorption frequency}\

\begin{deluxetable}{ccccc}
\tabletypesize{\scriptsize}
%\rotate
\tablecolumns{8}
\tablewidth{0pc}
\tablecaption{The constrained physical parameters of the merger ejecta for nonthermal emission.}
\tablehead{  \multicolumn{2}{c}{Assumptions} & \colhead{$E_{\rm ej}$ (erg)} & \colhead{$\beta_{\min}$}  & \colhead{$\gamma_{\max}$}}
\startdata
\multicolumn{2}{c}{$\epsilon_B\la10^{-2}$}&  $\ga2\times10^{50}$  &   &   \\
\multicolumn{2}{c}{$n\la10^{-1}\unit{cm^{-3}}$}  &   &  $\ga0.6$  &   \\
\multirow{2}*{$n\sim10^{-2}\unit{cm^{-3}}$, $\epsilon_B\sim10^{-5}-10^{-3}$} & $t_{\rm turn}\sim150$day  &  $(0.3-5)\times10^{51}$ &$0.7-0.8$   \\
 & $E_{\rm fast}\sim E_\gamma$ & & & $3.7 -7.2$
\enddata
\label{Tab:constraint}
\end{deluxetable}


\begin{thebibliography}{}
%\expandafter\ifx\csname natexlab\endcsname\relax\def\natexlab#1{#1}\fi

\bibitem[Abbott et al.(2017a)]{gw} Abbott, B.~P., Abbott, R., Abbott, T.~D., et al.\ 2017a, Physical Review Letters, 119, 161101
\bibitem[Abbott et al.(2017b)]{mm} Abbott, B.~P., Abbott, R., Abbott, T.~D., et al.\ 2017b, \apjl, 848, L12 %mm
\bibitem[Achterberg et al.(2001)]{pvalue3} Achterberg, A., Gallant, Y.~A., Kirk, J.~G., \& Guthmann, A.~W.\ 2001, \mnras, 328, 393

\bibitem[Alexander et al.(2017)]{2017ApJ...848L..21A} Alexander, K.~D., Berger, E., Fong, W., et al.\ 2017, \apjl, 848, L21 % radio, apjl
\bibitem[Alexander et al.(2018)]{ale18} Alexander, K.~D., Margutti, R., Blanchard, P.~K., et al.\ 2018, arXiv:1805.02870  %fading X-ray
\bibitem[Arcavi et al.(2017)]{arc17} Arcavi, I., Hosseinzadeh, G., Howell, D.~A., et al.\ 2017, \nat, 551, 64 %thermal optical

\bibitem[Barniol Duran(2014)]{bar14} Barniol Duran, R.\ 2014, \mnras, 442, 3147

\bibitem[Bednarz \& Ostrowski(1998)]{pvalue1} Bednarz, J., \& Ostrowski, M.\ 1998, Physical Review Letters, 80, 3911

\bibitem[Berger(2014)]{berger} Berger, E.\ 2014, \araa, 52, 43

\bibitem[Blandford \& McKee(1976)]{bm76} Blandford, R.~D., \& McKee, C.~F.\ 1976, Physics of Fluids, 19, 1130

\bibitem[Coulter et al.(2017)]{cou17}
Coulter, D.~A., Foley, R.~J., Kilpatrick, C.~D., et al.\ 2017, Science, 358, 1556

\bibitem[Dai et al.(1999)]{dai99} Dai, Z.~G., Huang, Y.~F., \& Lu, T.\ 1999, \apj, 520, 634

\bibitem[D'Avanzo et al.(2018)]{x-decreasing} D'Avanzo, P., Campana, S., Ghisellini, G., et al.\ 2018, arXiv:1801.06164  %XMM, x-ray decreasing; jet+isotropic, false reference
\bibitem[Dobie et al.(2018)]{radio-turn} Dobie, D., Kaplan, D.~L., Murphy, T., et al.\ 2018, arXiv:1803.06853 % radio turnover

\bibitem[Eichler et al.(1989)]{eichler} Eichler, D., Livio, M., Piran, T., \& Schramm, D.~N.\ 1989, \nat, 340, 126

\bibitem[Evans et al.(2017)]{evans17} Evans, P.~A., Cenko, S.~B., Kennea, J.~A., et al.\ 2017, Science, 358, 1565 %thermal, swift, science

\bibitem[Fong et al.(2015)]{fon15}
Fong, W., Berger, E., Margutti, R., \& Zauderer, B.~A.\ 2015, \apj, 815, 102 %short GRB density

\bibitem[Goldstein et al.(2017)]{GBM} Goldstein, A., Veres, P., Burns, E., et al.\ 2017, \apjl, 848, L14  % gamma-ray, GBM
\bibitem[Gottlieb et al.(2017)]{cocoon} Gottlieb, O., Nakar, E., Piran, T., \& Hotokezaka, K.\ 2017, arXiv:1710.05896 %cocoon

\bibitem[Haggard et al.(2017)]{2017ApJ...848L..25H} Haggard, D., Nynka, M., Ruan, J.~J., et al.\ 2017, \apjl, 848, L25  %X-ray Chandra, apjL
\bibitem[Haggard et al.(2018)]{latestX} Haggard, D., Nynka, M., Ruan, J.~J., Evans, P., \& Kalogera, V.\ 2018, The Astronomer's Telegram, 11242,

\bibitem[Hallinan et al.(2017)]{2017Sci...358.1579H} Hallinan, G., Corsi, A., Mooley, K.~P., et al.\ 2017, Science, 358, 1579  %radio Science

\bibitem[Hjorth et al.(2017)]{hjorth17} Hjorth, J., Levan, A.~J., Tanvir, N.~R., et al.\ 2017, \apjl, 848, L31

\bibitem[Hotokezaka et al.(2018)]{dyn_ej} Hotokezaka, K., Kiuchi, K., Shibata, M., Nakar, E., \& Piran, T.\ 2018, arXiv:1803.00599   %fast tail synchrotron

\bibitem[Keshet \& Waxman(2005)]{pvalue4} Keshet, U., \& Waxman, E.\ 2005, Physical Review Letters, 94, 111102

\bibitem[Kirk et al.(2000)]{pvalue2} Kirk, J.~G., Guthmann, A.~W., Gallant, Y.~A., \& Achterberg, A.\ 2000, \apj, 542, 235

\bibitem[Lazzati et al.(2017)]{lazzjet} Lazzati, D., Perna, R., Morsony, B.~J., et al.\ 2017, arXiv:1712.03237 %jet
\bibitem[Lemoine et al.(2013)]{lemoine13} Lemoine, M., Li, Z., \& Wang, X.-Y.\ 2013, \mnras, 435, 3009

\bibitem[Levan et al.(2017)]{levan17} Levan, A.~J., Lyman, J.~D., Tanvir, N.~R., et al.\ 2017, \apjl, 848, L28

\bibitem[Lyman et al.(2018)]{optical1} Lyman, J.~D., Lamb, G.~P., Levan, A.~J., et al.\ 2018, arXiv:1801.02669 %optical, one point; jet

\bibitem[Margutti et al.(2017)]{2017ApJ...848L..20M} Margutti, R., Berger, E., Fong, W., et al.\ 2017, \apjl, 848, L20  %X-ray
\bibitem[Margutti et al.(2018)]{margutti18} Margutti, R., Alexander, K.~D., Xie, X., et al.\ 2018, arXiv:1801.03531 %jet
\bibitem[Mooley et al.(2018)]{mooley17} Mooley, K.~P., Nakar, E., Hotokezaka, K., et al.\ 2018, \nat, 554, 207
\bibitem[Murguia-Berthier et al.(2017)]{mur17} 
Murguia-Berthier, A., Ramirez-Ruiz, E., Kilpatrick, C.~D., et al.\ 2017, \apjl, 848, L34 
%\bibitem[Nakar \& Piran(2017)]{nak17} Nakar, E., \& Piran, T.\ 2017, \apj, 834, 28
\bibitem[Nakar \& Piran(2018)]{np18} Nakar, E., \& Piran, T.\ 2018, arXiv:1801.09712

\bibitem[Nakar et al.(2018)]{nakar_whole} Nakar, E., Gottlieb, O., Piran, T., Kasliwal, M.~M., \& Hallinan, G.\ 2018, arXiv:1803.07595
\bibitem[Nynka et al.(2018)]{nyn18} Nynka, M., Ruan, J.~J., \& Haggard, D.\ 2018, arXiv:1805.04093
\bibitem[Pian et al.(2017)]{pian17} Pian, E., D'Avanzo, P., Benetti, S., et al.\ 2017, \nat, 551, 67 %thermal

\bibitem[Ruan et al.(2017)]{ruan17} Ruan, J.~J., Nynka, M., Haggard, D., Kalogera, V., \& Evans, P.\ 2017, arXiv:1712.02809 % steep X-ray spectrum at 16 day, but not in Margutti et al.
\bibitem[Resmi et al.(2018)]{resmi18} Resmi, L., Schulze, S., Ishwara Chandra, C.~H., et al.\ 2018, arXiv:1803.02768 %low freq radio, 1.4Ghz

\bibitem[Santana et al.(2014)]{san14} Santana, R., Barniol Duran, R., \& Kumar, P.\ 2014, \apj, 785, 29 %magnetic field for GRB

\bibitem[Sari et al.(1998)]{sari98} Sari, R., Piran, T., \& Narayan, R.\ 1998, \apjl, 497, L17

\bibitem[Savchenko et al.(2017)]{2017ApJ...848L..15S} Savchenko, V., Ferrigno, C., Kuulkers, E., et al.\ 2017, \apjl, 848, L15 % gamma-ray, integral


\bibitem[Troja et al.(2017)]{2017Natur.551...71T} Troja, E., Piro, L., van Eerten, H., et al.\ 2017, \nat, 551, 71  %X-ray Nature

\bibitem[Troja et al.(2018)]{troja18} Troja, E., et al.\ 2018, arXiv:1801.06516  %cocoon + jet
%\bibitem[Troja \& Piro(2018)]{latestX2} Troja, E., \& Piro, L.\ 2018, The Astronomer's Telegram, 11245,

\bibitem[Waxman et al.(2017)]{waxman17} Waxman, E., Ofek, E., Kushnir, D., \& Gal-Yam, A.\ 2017, arXiv:1711.09638

\bibitem[Wijers \& Galama(1999)]{wijers99} Wijers, R.~A.~M.~J., \& Galama, T.~J.\ 1999, \apj, 523, 177

\end{thebibliography}
\end{document}